\documentclass[conference]{IEEEtran}
\usepackage{cite}
\usepackage{amsmath,amssymb,amsfonts}
\usepackage{algorithmic}
\usepackage{graphicx}
\usepackage{textcomp}
\usepackage{xcolor}
\usepackage[colorlinks,urlcolor=blue,citecolor=blue,linkcolor=blue]{hyperref}
\usepackage{mathtools}
\usepackage{tikz}
\usepackage{caption}
\usepackage{subcaption}
\usetikzlibrary{positioning}
\DeclarePairedDelimiter\ceil{\lceil}{\rceil}
\DeclarePairedDelimiter\floor{\lfloor}{\rfloor}

\usepackage[a4paper, total={184mm,239mm}]{geometry}
\usepackage{multirow} 
\newcommand{\SR}{\text{SR}}
\newcommand{\tr}{\text{tr}}

\begin{document}

\title{A Stochastic Rounding-Enabled Low-Precision Floating-Point MAC for DNN Training}



\author{\IEEEauthorblockN{Sami Ben Ali,
Silviu-Ioan Filip, Olivier Sentieys}
\IEEEauthorblockA{Univ. Rennes, Inria\\
Rennes, France\\
\{sami.ben-ali, silviu.filip, olivier.sentieys\}@inria.fr}}
\maketitle

\maketitle

\begin{abstract}
  Training Deep Neural Networks (DNNs) can be computationally demanding, particularly when dealing with large models. Recent work has aimed to mitigate this computational challenge by introducing 8-bit floating-point (FP8) formats for multiplication. However, accumulations are still done in either half (16-bit) or single (32-bit) precision arithmetic. In this paper, we investigate lowering accumulator word length while maintaining the same model accuracy. 
  We present a multiply-accumulate (MAC) unit with FP8 multiplier inputs and FP12 accumulations, which leverages an optimized stochastic rounding (SR) implementation to mitigate swamping errors that commonly arise during low precision accumulations. We investigate the hardware implications and accuracy impact associated with varying the number of random bits used for rounding operations. We additionally attempt to reduce MAC area and power by proposing a new scheme to support SR in floating-point MAC and by removing support for subnormal values. Our optimized \textit{eager}  SR unit significantly reduces delay and area when compared to a classic \textit{lazy} SR design. Moreover, when compared to MACs utilizing single- or half-precision adders, our design showcases notable savings in all metrics. Furthermore, our approach consistently maintains near baseline accuracy across a diverse range of computer vision tasks, making it a promising alternative for low-precision DNN training.
\end{abstract}


\section{Introduction}
Deep Neural Networks (DNNs) have emerged as a cornerstone of modern artificial intelligence, enabling groundbreaking advancements in areas such as image recognition, natural language processing, and autonomous systems. The success of DNNs, however, comes at a significant cost. Training modern networks requires substantial computational resources, making it an arduous and resource-intensive process. As the scale and complexity of DNN models continue to grow, the demand for efficient training methods becomes increasingly urgent.

In response to the computational demands of DNN training, recent research has explored the use of lower-precision arithmetic. By reducing the bit-width of numerical representations, such as using 8-bit floating-point (FP8) formats, significant computational savings can be achieved. However, while much attention has been given to optimizing the precision of multiplicative operations, accumulative operations in DNNs, such as summations and weight updates, often remain in higher-precision formats like half-precision (16-bit) or single-precision (32-bit). This disparity in precision can lead to inefficiencies in hardware utilization and energy consumption.

In this paper, we attempt to reduce the precision of accumulations in forward (FWD) and backward (BWD) matrix multiply operations during training while preserving the model's accuracy. Our primary focus is  the design of a dedicated multiply-accumulate (MAC) unit optimized for efficiently handling FP8 multiplication inputs and FP12 accumulations. Recognizing that low-precision arithmetic can limit accuracy, we use stochastic rounding to mitigate the accuracy loss.

The paper is organized as follows. 
Section \ref{sec:background} presents some background on stochastic rounding (SR) as well as related work on low-precision DNN training and SR hardware.
Section \ref{sec:contrib} details our main contribution on an SR-enabled floating-point MAC unit design. 
Finally, Section \ref{sec:results} presents some results on low-precision training showing that our approach consistently maintains near baseline accuracy across a diverse range of image classification tasks.

\section{Background and Related Work}
\label{sec:background}
Rounding errors pose a prominent challenge within floating-point arithmetic. While DNN training can exhibit some resilience to these errors, their cumulative impact may ultimately result in substantial information loss. Various research efforts~\cite{wang2018training,zamirai2020revisiting} study the potential benefits of employing stochastic rounding as a strategy to mitigate the effects of rounding errors in this context. SR is particularly effective
against stagnation~\cite{blanchard2020class}, a frequent occurance when computing the sum of a large number of terms with small magnitude and a large forward error is produced.

\subsection{Stochastic Rounding}\label{sec:sota_sr}

As opposed to deterministic rounding schemes, SR randomly maps a real value $x$ to one of the two closest values in the number system being used. The probability of choosing either one is given by $1$ minus its relative distance to $x$. More formally, for a finite set $\mathbb{F}\subseteq\mathbb{R}$, we denote the two rounding candidates
\[
  \floor*{x}=\max\{y\in\mathbb{F}|y\leqslant x\} \text{ and } \ceil*{x}=\min\{y\in\mathbb{F}|y\geqslant x\}.
\]
If $x\notin \mathbb{F}$, we have that (``w.p'' stands for ``with probability'')
\[
  \SR(x) = \left\{
  \begin{array}{ll}
    \ceil*{x},  & \mbox{w.p.} \ q(x) = \frac{x - \floor*{x} }{\ceil*{x} - \floor*{x}}, \\
    \floor*{x}, & \mbox{w.p.} \  1-q(x).
  \end{array}
  \right.
\]

In case $\mathbb{F}$ is a normalized $p$ binary digit precision floating-point system with maximum exponent $e_{\max}$, machine epsilon $\varepsilon:=2^{1-p}$, and if $x\in\mathbb{F}$, then it can be written as $x=(-1)^{s_x}\cdot 2^{e_x}\cdot m_x$, where $s_x\in\{0,1\}$ is the sign bit, $e_x$ is an integer exponent between $e_{\min}:=1-e_{\max}$ and $e_{\max}$ inclusive, and $m_x\in[0,2)$ can be represented exactly using $p$ binary digits, $p-1$ of which are stored explicitly. The case $m_x<1$ corresponds to subnormal values and only occurs when $e_x=e_{\min}$.

If $x\in\mathbb{R}$ is written in normalized form as $(-1)^{s_x}\cdot 2^{e_x}\cdot m_x$, but this time with real-valued $m_x\in[0,2)$, and we denote by
$\tr(m_x):=2^{1-p}\floor*{2^{p-1}m_x}$ its $p$-digit truncation, then $\SR(x)$ can be written as~\cite[Eq.~(4.1)]{fasi2021algorithms}
\begin{equation}\label{eq:sr_tr}
  \SR(x) = \left\{
  \begin{array}{ll}
    (-1)^{s_x}\cdot 2^{e_x} \cdot \tr(m_x),                          & \mbox{w.p.} \ 1-\varepsilon_x, \\
    (-1)^{s_x}\cdot 2^{e_x} \cdot \left(\tr(m_x)+\varepsilon\right), & \mbox{w.p.} \  \varepsilon_x,
  \end{array}
  \right.
\end{equation}
where $\varepsilon_x:=(m_x-\tr(m_x))/\varepsilon\in[0,1)$ and $|x|$ is between the smallest and largest representable positive values in $\mathbb{F}$.

From an implementation point of view, it makes more sense to consider the equivalent formulation~\cite[Eq.~(4.4)]{fasi2021algorithms} of~\eqref{eq:sr_tr}
\begin{equation}\label{eq:sr_rng}
  \SR(x) = \left\{
  \begin{array}{ll}
    (-1)^{s_x}\cdot 2^{e_x} \cdot \tr(m_x),                          & X\geqslant \varepsilon_x, \\
    (-1)^{s_x}\cdot 2^{e_x} \cdot \left(\tr(m_x)+\varepsilon\right), & X < \varepsilon_x,
  \end{array}
  \right.
\end{equation}
where $X\sim\mathcal{U}_{[0,1)}$ is a random variable uniformly distributed over $[0,1)$. Definition~\eqref{eq:sr_rng} holds true in the discrete case as well~\cite[Sec.~4]{fasi2021algorithms}: if $X$ is generated on $r$ bits, then $x$ will be rounded up in $2^r \varepsilon_x$ cases out of $2^r$. What value of $r$ is best in an SR implementation is an open question. It is certain that a lower value of $r$ leads to a cheaper hardware implementation, but it will impact the accuracy benefit that SR potentially brings.

\begin{figure}[t]
  \centering
  \includegraphics[width=0.4\textwidth]{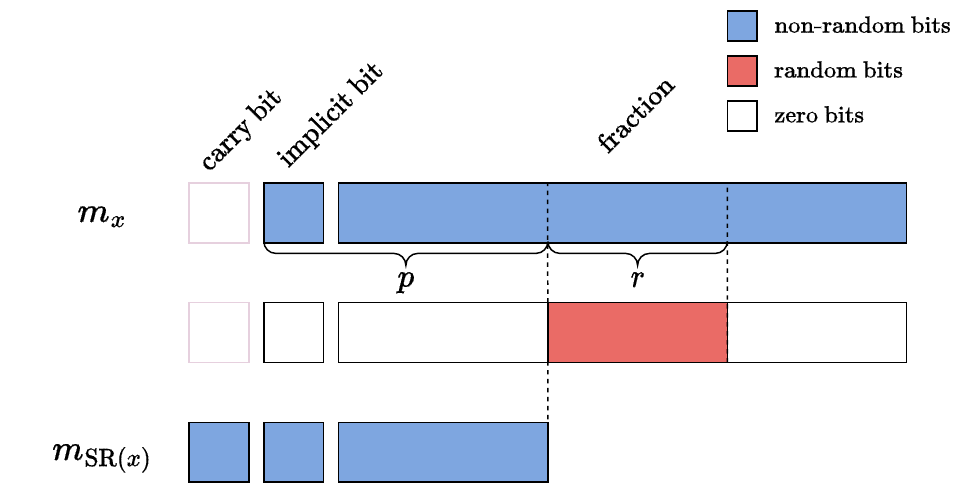}
  \caption{Bit alignment in algorithms for sum-based stochastic rounding implementations. The random bits are added to the significand $m_x$, followed by its truncation.}
  \label{fig:sr_basic_implem}
  \vspace{-10pt}
\end{figure}

Most hardware and software implementations of SR (see~\cite[Sec.~7.3--7.4]{croci2022stochastic}) do not directly use the comparison suggested by~\eqref{eq:sr_rng}, but instead take the equivalent route of adding bits from a random source to the part of the number that will be truncated. If a carry out occurs after this addition, the rounded up value will be the result. A pictorial representation of this process is given in Fig.~\ref{fig:sr_basic_implem} (adapted from~\cite[Fig.~2]{croci2022stochastic}). Exceptional value handling can be done as suggested in~\cite[Sec.~7.2]{croci2022stochastic}.

\subsection{Low Precision DNN Training}\label{sec:sota_lptrain}
The FWD and BWD passes in deep learning models are frequently implemented as General Matrix Multiplications (GEMM), which constitute the most computationally demanding aspect of a training algorithm. Recent research has introduced the adoption of FP8 formats for multiplicative operations within GEMM procedures~\cite{micikevicius2022fp8,sun2019hybrid,cambier2020shifted,mellempudi2019mixed,tatsumi2022mixing}. Although quantizing inputs into 8-bit formats substantially reduces memory access time for transferring operands, it is essential to note that in most related work, accumulations are still conducted using 16-bit or 32-bit floating-point formats, jeopardizing overall gains. Moreover, quantizing the inputs leads to potential accuracy degradation due to the limited representation range of FP8 formats. To address this issue, some approaches employ loss scaling to mitigate accuracy loss~\cite{mellempudi2019mixed,tatsumi2022mixing,micikevicius2017mixed}. Others use distinct FP8 formats for the FWD and BWD passes, while introducing roundoff schemes to minimize memory access overhead~\cite{sun2019hybrid}.

Additional work explores the use of fixed-point accumulators in conjunction with FP8 multipliers to reduce the area overhead associated with FP16 accumulators~\cite{tatsumi2022mixing}.
Beyond FP8 formats, alternative formats for DNN training, such as BFloat16~\cite{osorio2022bf16,kalamkar2019study} and block floating-point formats~\cite{zhang2022fast}, have also been investigated. Approaches like~\cite{sun2020ultra} lower precision further by employing 4-bit integer arithmetic, complemented with gradient scaling to minimize accuracy loss, but for small DNN models.

Several studies utilize SR in low-precision DNN training. However, when it comes to hardware implementations, a predominant focus has been on integrating SR with integer arithmetic~\cite{gupta2015deep,zhang2022fast,chang2023esru} due to its simplicity and low hardware cost. While some work explores its usage in conjunction with floating-point arithmetic~\cite{wang2018training,mellempudi2019mixed}, the primary focus is evaluating accuracy gains rather than delving into hardware specifics.

More recent research focuses on optimizing hardware for stochastic rounding units. Within this context, an efficient stochastic rounding unit (SRU) was introduced, relying on a reduced bitstream for random values~\cite{chang2023esru}. Furthermore, another work~\cite{yuan2022you} seeks to minimize the SRU's footprint by utilizing entropy generated during model training instead of hardware-based random generators. While such hardware optimizations predominantly adhere to integer arithmetic, our work explores SR hardware implementation within floating-point MAC units, and its application to the training of DNN models.

\section{SR-Enabled Floating-Point MAC Unit Design}
\label{sec:contrib}
\begin{figure}[t]
  \centering
\includegraphics[width=0.25\textwidth]{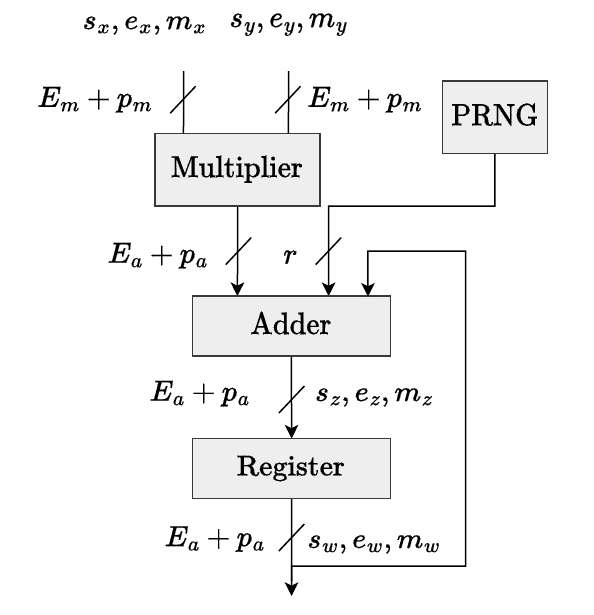} 
  \caption{Overview of our floating-point MAC unit designs with SR support. The multiplier results are exact, with rounding only being done in the adder part.}
  \label{fig:mac}
  \vspace{-10pt}
\end{figure}

\begin{figure*}[t!]
  \centering
      \vspace{-0.5cm}

  \begin{subfigure}[t]{0.35\textwidth} 
    \centering
    \includegraphics[width=\textwidth]{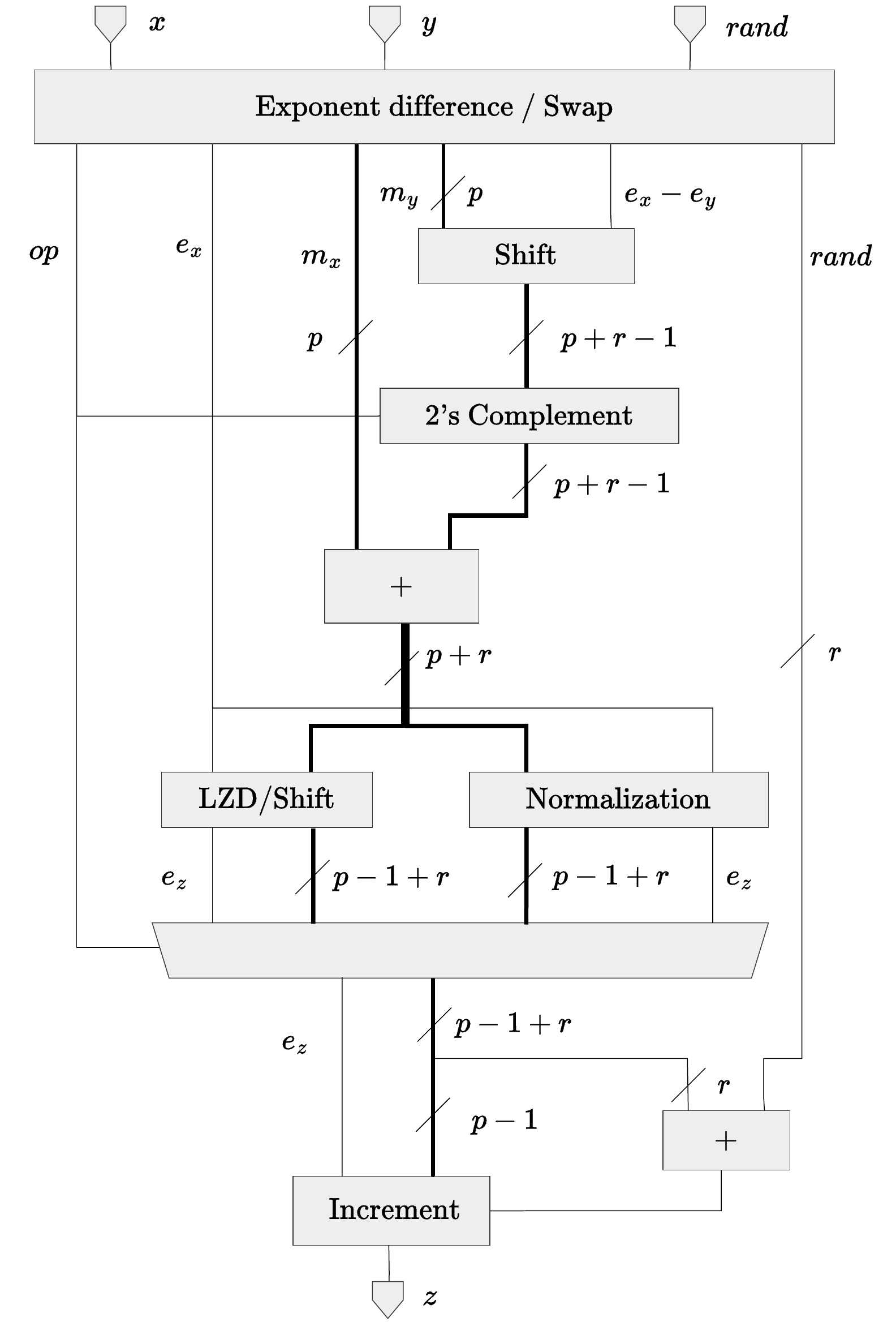}
    \caption{FP Adder with \emph{lazy} SR}
    \label{fig:sr_add_lazy}
  \end{subfigure}
  \hfill 
  \begin{subfigure}[t]{0.577\textwidth} 
    \centering
    \includegraphics[width=\textwidth]{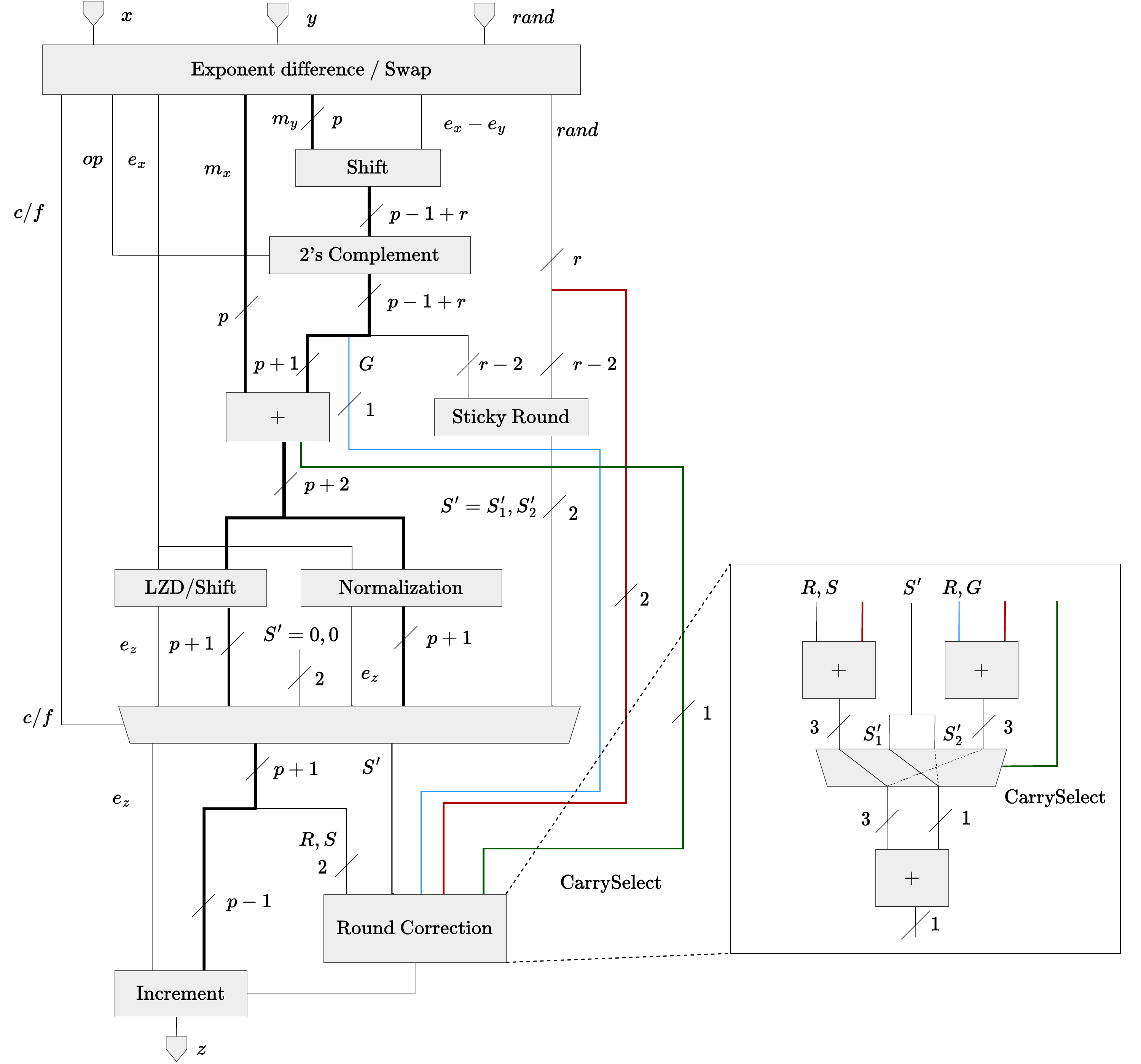}
    \caption{FP Adder with \emph{eager} SR}
    \label{fig:sr_add_eager}
  \end{subfigure}

  \caption{Two possible implementations of a SR-enabled floating-point adder. The first (a) does stochastic rounding in a lazy fashion late in the addition process, after normalization, whereas the second (b) starts the process early, with the caveat that a small rounding correction is required towards the end of the process.}
  \label{fig:sr_add_main}
\end{figure*}

Our MAC unit designs are composed of three major parts, as summarized in Figure~\ref{fig:mac}.
\paragraph{\textbf{Multiplier}} This is an exact variant that computes the product of two $p_m$-bit precision values with $E_m$ exponent bits as a $p_a:=2p_m$-bit precision result with $E_a:=E_m+1$ exponent bits. Taking this full result eliminates the need for rounding that would otherwise consume extra logic. For example, our reference FP8 design with E$5$M$2$ multiplier inputs will output FP12 E$6$M$5$ results.
\paragraph{\textbf{Accumulator}} The multiplier feeds into a $p_a$-bit precision adder, with its other input being the previous adder output. An in-depth presentation is left to Sec.~\ref{sec:lazy_sr} and Sec.~\ref{sec:eager_sr}.

\paragraph{\textbf{Random Number Generator}}The design is completed by a $r$-bit pseudo-random number generator (PRNG) that operates in parallel and asynchronously with the multiplier and whose output will be used in the SR operation of the adder. It is based on a Galois linear feedback shift register (LFSR). 



\subsection{Standard SR-Enabled Floating-Point Adder}\label{sec:lazy_sr}
The most involved part in the unit is the adder and the integration of SR. To highlight what happens, we first consider the case where round-to-nearest-even (RN) is performed and then present the changes needed to support SR. Our discussion partially follows~\cite[Sec.~7.3]{croci2022stochastic}, using a dual-path floating-point adder design\footnote{Our design is a variation on a standard dual-path floating-point adder~\cite[Fig.~8.11]{MullerEtAl2018}. The main difference is that we use one integer adder/subtractor circuit for computing the (un-normalized) significand of the result, as opposed to the vanilla design that has a different circuit for each path. This saves some area at the expense of latency. The ideas for our SR-enabled units should be applicable to other adder designs, such as the aforementioned standard one.}. For simplicity, we assume $s_x=s_y=0$ to avoid sign interactions, which might transform the addition into a subtraction. The analysis carries over to effective subtraction.

Let $x=2^{e_x}m_x$ and $y=2^{e_y}m_y$ be the precision $p$ normalized inputs in the
adder (\emph{i.e.,}~$m_x=\text{tr}(m_x)$ and $m_y=\text{tr}(m_y)$). The basic addition algorithm can be summarized as follows:
\begin{itemize}
  \item[(i)] \textbf{reordering and swap}: if $e_y>e_x$, swap $x$ and $y$.
  \item[(ii)] \textbf{significand alignment}: compute $m_y 2^{-(e_x-e_y)}$ (\emph{i.e.,}~shift $m_y$ to the right $e_x-e_y$ places).
  \item[(iii)] \textbf{significand addition}: compute $m_t=m_x + m_y2^{-(e_x-e_y)}$.  The situation $e_x - e_y>1$ is handled by the far ($f$) path, whereas $e_x - e_y\leqslant 1$ is handled by the close ($c$) path.
  \item[(iv)] \textbf{normalization}: since $m_t\in[0,2^{p+1})$ we might have to shift $m_z$ to the right by one bit (if $m_t\geqslant 2^p$) and increment $e_z$. A shift left is possible with effective subtraction, handled with a leading zero detector (LZD) and shifter.
  \item[(v)] \textbf{rounding}: the significand of the rounded sum $m_z$ is computed by rounding the normalized exact sum $m_t$. In case of RN, we need bits at positions $p+1$, $p+2$ in the significand (the guard and round bits), and the sticky bit (logical OR of all the bits after the $(p+2)$nd). The sticky bit can be computed during (ii)--(iii).
\end{itemize}

A straightforward implementation of SR in the floating-point adder case would simply amount to replacing the sticky bit computation (and that of the guard and round bits) with adding $r$ randomly generated bits to the normalized result significand starting at position $p+1$, like in our discussion from Sec.~\ref{sec:sota_sr}. Figure~\ref{fig:sr_add_lazy} summarizes this, and it seems that most hardware SR implementations follow this approach~\cite[Sec.~7.3]{croci2022stochastic}\footnote{There is a patent from IBM on floating-point addition with SR~\cite{bradbury2019stochastic}, but it is not clear based on the given description if the rounding operation is performed after normalization or not.}. Effective $+/-$ (values of $s_x$ and $s_y$) is handled with the $op$ flag.


\subsection{Reduced-Latency SR-Enabled Floating-Point Adder}\label{sec:eager_sr}
The rounding is deferred until the result has been normalized. Such a~\emph{lazy} execution impacts the critical path, and one can think that latency can be improved if rounding is started earlier, in an~\emph{eager} fashion. This is the premise of our optimized eager SR-enabled floating-point adder whose schema is given in Figure~\ref{fig:sr_add_eager}. The rounding operation starts immediately after the significand alignment (ii), where we tentatively add the $r-2$ least significant bits (LSBs) of the generated random bitstring to the $r-2$ bits from $y$'s shifted significand (the \textbf{Sticky Round} block), starting at position $p+3$. The $p+1$ most significant bits (MSBs) of the shifted value of $y$ are used in computing the addition, whereas the two most significant bits of the \textbf{Sticky Round} output, denoted with $S'=S'_1S'_2$ are potentially used in a \textbf{Round Correction} stage that occurs after the normalization of the addition result. For our implementation, the \textbf{Normalization} block relies on a carry-dependent $1$-bit left shift. In the event of a carry during addition, the new carry bit becomes the updated implicit bit while the exponent is incremented. Conversely, when no carry occurs, the result requires a left shift by one bit. Additionally, the implicit bit will no longer be needed. Therefore, the block outputs a denormalized value.

\begin{figure}[t]
  \begin{subfigure}{0.5\textwidth}  
    \centering
    \includegraphics[width=0.65\linewidth]{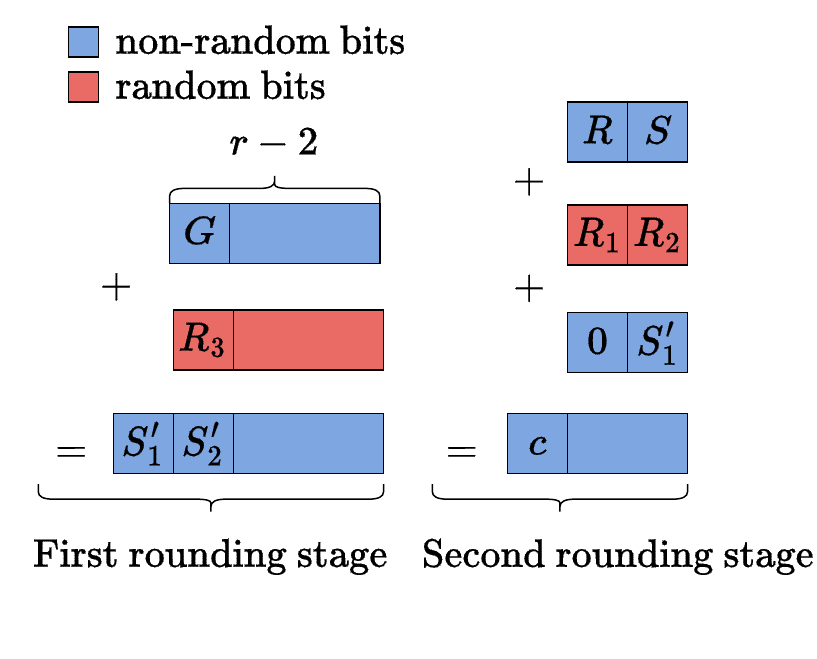}
    \vspace{-0.3cm}
    \caption{Eager SR when no normalization happens}
    \label{fig:subfig51}
  \end{subfigure}
  \begin{subfigure}{0.5\textwidth}
    \centering
    \includegraphics[width=0.65\linewidth]{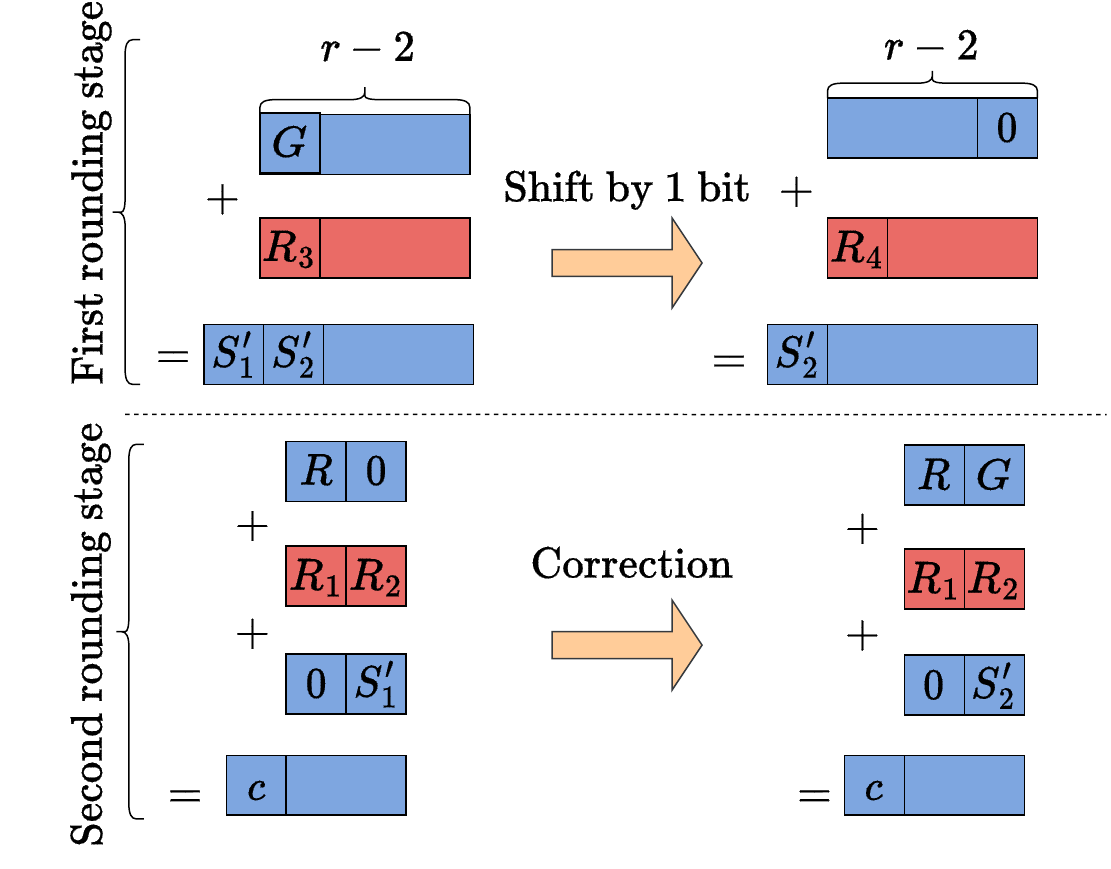}
    \caption{Eager SR when normalization happens}
    \label{fig:subfig52}
  \end{subfigure}
  \caption{Detailed flow of eager SR Round Correction operation in the two possible normalization cases.}
  \label{fig:main}
  \vspace{-10pt}
\end{figure}

There are two main cases to consider:
\begin{itemize}
  \item[(a)] \textbf{When no normalization takes place}:
    In this case, the addition result remains unshifted, and employing the eager design produces an identical outcome to calculating the rounding carry bit $c$ as with the lazy implementation. The sole distinction is that, in the eager approach, the rounding addition unfolds across two stages, as shown in Figure~\ref{fig:subfig51}.

  \item[(b)] \textbf{When normalization takes place}:
    In this case, the addition result undergoes a leftward shift by $1$ bit, the normalized value consistently retains an LSB value of $0$ instead of aligning with the $G$ bit value (as it would have been shifted into that position). To address this inherent discrepancy, we take corrective action by substituting the LSB of the normalized result with the $G$ bit, as shown in Figure~\ref{fig:subfig52}. As this normalization shift also influences the effect of the \textbf{Sticky Round} output, the carry from the first stage becomes $S’_2$ instead of $S’_1$. Through this correction, we achieve results that mirror those of the classic implementation.
\end{itemize}


To further validate the eager SR adder design, we also conduct brute-force testing using a vast array of $10000$ input pairs covering all the possible execution traces in the adder architecture. For every combination of input values $x$ and $y$, we employ $1000$ random integers and we calculate the probability of rounding occurrence accurately. We verify that, for each input configuration, the calculated probability aligns with the stochastic rounding definition outlined in Sec.~\ref{sec:sota_sr}. This testing process further confirms the precision and adherence of the implementation to the specified stochastic rounding criteria.

\subsection{Hardware Synthesis Results}\label{sec:hw_results}






\subsubsection{Experimental Setup}

We developed an RTL implementation for the various MAC units. The objective is to examine the tradeoffs concerning power, area, and delay under different configurations and bit widths. To this end, we conducted RTL synthesis using Synopsys Design Vision 2019.03, in conjunction with an FDSOI 28nm technology. We also performed FPGA synthesis using Vivado 2022.1, targeting the Virtex UltraScale+ VU9P chip from Xilinx.
During our experiments, we relax timing constraints and optimize design area.

\begin{figure*}[thbp]
  \centering
    \vspace{-0.3cm}

  \begin{subfigure}{0.31\textwidth} 
    \centering
    \includegraphics[width=\linewidth]{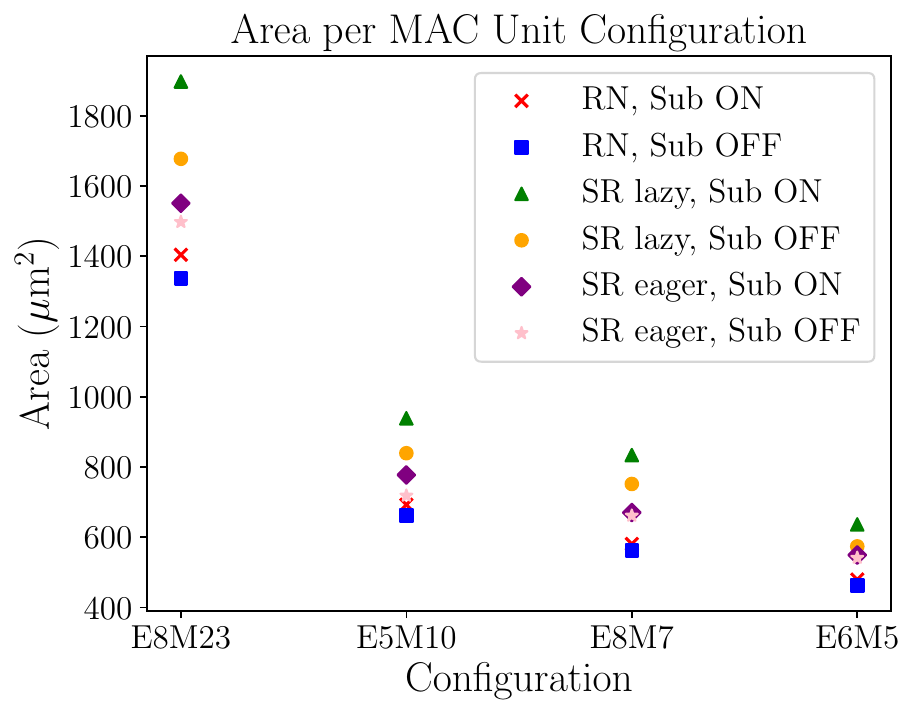}
    \caption{Area per configuration}
    \label{fig:area conf}
  \end{subfigure}
  \begin{subfigure}{0.295\textwidth} 
    \centering
    \includegraphics[width=\linewidth]{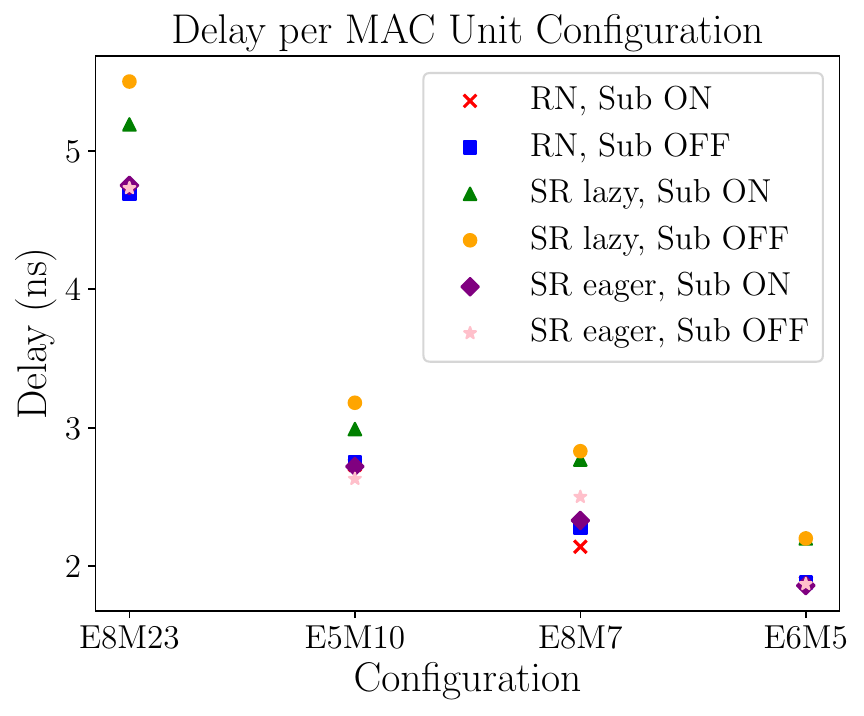}
    \caption{Delay per configuration}
    \label{fig:delay conf}
  \end{subfigure}
  \begin{subfigure}{0.30\textwidth} 
    \centering
    \includegraphics[width=\linewidth]{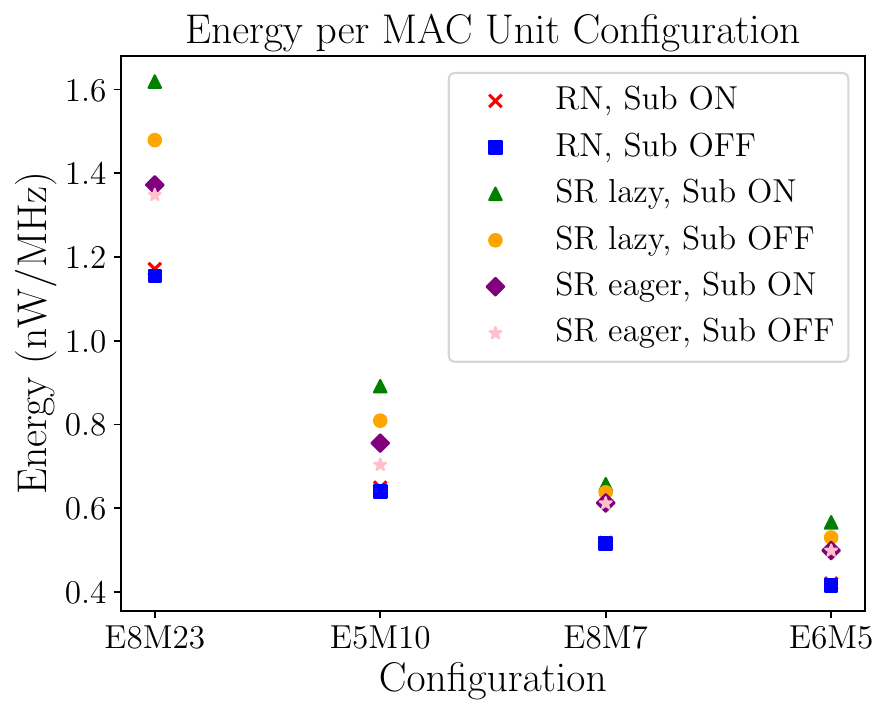}
    \caption{Energy per configuration}
    \label{fig:energy conf}
  \end{subfigure}

  \caption{Hardware cost for different floating-point MAC configurations.}
  \label{fig:overead_fig}
  \vspace{-10pt}
\end{figure*}

\subsubsection{Synthesis results}
The synthesis results are presented in Figure \ref{fig:overead_fig} and Table \ref{tab:Hardware configs}, which report the Energy, Area and Delay of various MAC configurations (RN, SR lazy, SR eager, with and without subnormals support), and for various adder bit widths (FP32, FP16, BFloat16, E6M5). FP32 corresponds to E8M23, FP16 to E5M10, Bfloat16 to E8M7, and E6M5 is our proposed 12-bit low-precision accumulator format.

Under identical configurations and bit settings, our optimized (eager) SR MAC design consistently outperforms the classic (lazy) SR implementation across all test metrics. The area gain is mainly due to having larger LZD and Normalization blocks in the classic case ($p+r$ versus $p+2$). This trend carries over when excluding support for subnormal values. In these assessments we set $r$ to $p + 3$. This selection is made to align with the IEEE-754 definition of RN, ensuring consistency in the number of bits retained after shifting. In the subsequent section, we delve deeper into the impact of the value of $r$ on both hardware performance and model accuracy.

\begin{table}[t]
  \centering
  \caption{Hardware cost for different FP adder configurations.}
  \label{tab:Hardware configs}
  \begin{tabular}{cccccccc}
    \hline
    Configuration                     & E & M  & $r$  & Energy     & Area        & Delay \\
    &   &    &    & ($nW/MHz$) & ($\mu m^2$) & (ns)  \\
    \hline
    \multirow{4}{*}{RN W/ Sub}        & 8 & 23 & 0  & \textbf{1.17}       & \textbf{1404.01}     & \textbf{4.71}  \\
    & 5 & 10 & 0  & \textbf{0.65}       & \textbf{692.62}      & \textbf{2.73}  \\
    & 8 & 7  & 0  & 0.52       & 581.05      & 2.14  \\
    & 6 & 5  & 0  & 0.42       & 479.81      & 1.88  \\
    \hline
    \multirow{4}{*}{RN W/O Sub}       & 8 & 23 & 0  & 1.15       & 1337.42     & 4.69  \\
    & 5 & 10 & 0  & 0.64       & 662.43      & 2.75  \\
    & 8 & 7  & 0  & 0.52       & 562.44      & 2.28  \\
    & 6 & 5  & 0  & 0.42       & 462.67      & 1.88  \\\hline
    \multirow{4}{*}{SR lazy W/ Sub}   & 8 & 23 & 27 & 1.62       & 1897.36     & 5.19  \\
    & 5 & 10 & 14 & 0.89       & 938.73      & 2.99  \\
    & 8 & 7  & 11 & 0.66       & 833.84      & 2.77  \\
   & 6 & 5  & 9  & 0.57       & 636.64      & 2.20  \\\hline
    \multirow{4}{*}{SR lazy W/O Sub}  & 8 & 23 & 27 & 1.48       & 1677.37     & 5.50  \\
    & 5 & 10 & 14 & 0.81       & 839.34      & 3.18  \\
    & 8 & 7  & 11 & 0.64       & 751.74      & 2.83  \\
    & 6 & 5  & 9  & \textbf{0.57}       & \textbf{615.10}      & \textbf{2.05}  \\  \hline
    \multirow{4}{*}{SR eager W/ Sub}  & 8 & 23 & 27 & 1.37       & 1550.89     & 4.75  \\
    & 5 & 10 & 14 & 0.76       & 777.48      & 2.72  \\
    & 8 & 7  & 11 & 0.61       & 670.41      & 2.33  \\
    & 6 & 5  & 9  & 0.50       & 549.49      & 1.87  \\\hline
    \multirow{4}{*}{SR eager W/O Sub} & 8 & 23 & 27 & 1.35       & 1497.52     & 4.73  \\
    & 5 & 10 & 14 & 0.70       & 718.41      & 2.63  \\
    & 8 & 7  & 11 & 0.61       & 661.54      & 2.50  \\
    & 6 & 5  & 9  & \textbf{0.51}       & \textbf{558.63}      & \textbf{1.87}  \\
    \hline
  \end{tabular}
\end{table}



The results from Table \ref{tab:res fpga} confirm that the eager MAC design still offers area and timing gains within an FPGA setup.

\section{Low-Precision Training Results}
\label{sec:results}




\begin{table}[tbp]
  \centering
  \caption{FPGA implementation results for FP adder designs.}
  \label{tab:res fpga}
  \begin{tabular}{lcccccc}
    \hline
    Configuration & E & M  & $r$  & LUT & FF & Delay (ns)\\
    \hline
    RN W/ Sub     & 5 & 10 & -  & 302   & 49& 8.30   \\
    RN W/O Sub     & 5 & 10  & -  & 301 & 49 & 8.29  \\
    SR lazy W/O Sub    & 6 & 5  & 13 & 344 & 59 & 8.76   \\
    SR eager W/O Sub    & 6 & 5  & 13 & 251 & 59 & 8.04   \\
    \hline
  \end{tabular}
  \vspace{-10pt}
\end{table}

As previously stated, our aim is to deploy SR-enabled MAC units suited for DNN training scenarios and explore their impact in terms of hardware complexity and model accuracy. We focus in particular on MAC units with FP8 multipliers and FP12 accumulators, since this configuration seems best suited for the models tested and because FP8 arithmetic is gaining traction in terms of commercial hardware support~\cite{micikevicius2022fp8}. To this end, we investigate two critical factors: the number of random bits $r$ used for the rounding and the impact of using subnormal encodings (which add some overhead to the hardware). 

In the following, all GEMM operations during training (FWD and BWD passes) are performed using low-precision MAC units. To examine the effects of these precision choices, we use a PyTorch software-based bit-accurate emulation flow of the MAC units. It uses custom CUDA kernels to expedite the training simulation runs to GPUs.

\subsection{Training settings}

We use stochastic gradient descent with a momentum coefficient of $0.9$ across all our experiments. The initial learning rates and weight decay were configured to $0.1/0.01$ and $0.0001/0.0005$ for training ResNet-20 and VGG16 models, respectively. We also use a batch size of $128$ for both models. To modulate the learning rate throughout training, we employed a cosine annealing scheduler. The ResNet-20 and VGG16 models were trained for $165$ and $200$ epochs, respectively. ResNet-50 training tests were ran for $100$ epochs with a $0.01$ initial learning rate and batch size of $16$. Additionally, a dynamic loss scaling technique~\cite{micikevicius2017mixed} was applied to all experiments, using an initial scaling factor of $1024$.

\subsection{Impact of subnormals and number of random bits}

To investigate the impact of the number of random bits on training accuracy, we opted for the FP12 format (E6M5). We started with a ResNet20 model and the CIFAR10 dataset. As is visible in Table~\ref{tab:res20 acc}, using only $r=4$ bits for the random values yielded notably inferior accuracy results. However, as we progressively increase the number of random bits, accuracy consistently improves. With 13 bits, we achieved accuracy levels close to the baseline with and without subnormal support\footnote{When we do not have subnormal support, values in the subnormal range are treated as zero.}.

\begin{table}[t]
  \centering
  \caption{Impact of number format (E, M) and random number bits $r$ on accuracy when training ResNet20 on CIFAR10.}
  \label{tab:res20 acc}
  \begin{tabular}{lcccc}
    \hline
    Configuration & E & M  & $r$  & Accuracy (\%)\\
    \hline
    FP32 Baseline & 8 & 23 & -  & \textbf{91.47}    \\
    RN W/ Sub     & 5 & 10 & -  & 91.1     \\
    RN W/ Sub     & 8 & 7 & -  & 88.79     \\
    RN W/ Sub     & 6 & 5  & -  & 83.03    \\
    SR W/ Sub     & 6 & 5  & 4  & 43.11    \\
    SR W/ Sub     & 6 & 5  & 9 & 89.34    \\
    SR W/ Sub     & 6 & 5  & 11 & 90.7     \\
    SR W/ Sub     & 6 & 5  & 13 & \textbf{91.39}    \\
    SR W/O Sub    & 6 & 5  & 11 & 90.67    \\
    SR W/O Sub    & 6 & 5  & 13 & \textbf{91.39}    \\
    \hline
  \end{tabular}
\end{table}

\begin{table}[t]
  \centering
  \caption{Impact of number format (E, M) and number of random bits $r$ on accuracy for VGG16 and ResNet50 models.}
  \label{tab:large model acc}
  \begin{tabular}{lcccccc}
    \hline
    Model/Dataset                       & Configuration & E & M  & $r$  & Accuracy (\%) \\
    \hline
    \multirow{3}{*}{VGG16/CIFAR10}      & FP32 Baseline & 8 & 23 & -  & 93.46    \\
        & RN W/ Sub     & 5 & 10 & -  & 93.06    \\
        & SR W/O Sub    & 6 & 5  & 13 & 93.11    \\ \hline
    \multirow{3}{*}{ResNet50/Imagewoof} & FP32 Baseline & 8 & 23 & -  & 80.94    \\
     & RN W/ Sub     & 5 & 10 & -  & 80.3     \\
    & SR W/O Sub    & 6 & 5  & 13 & 80.33    \\
    \hline
  \end{tabular}
  \vspace{-10pt}
\end{table}

We also ran tests using larger VGG16 and ResNet50 models. For VGG16, we utilized the CIFAR10 dataset, while the ResNet50 model was trained on a more challenging dataset, Imagewoof, a subset of ImageNet. As shown in Table~\ref{tab:large model acc}, FP12 with $13$ random bits and without subnormals outperforms the FP16 accumulator while incurring minimal accuracy degradation for both models.
RN results with sub 16-bit formats and without subnormal support are not reported since the accuracy degradation becomes significant in these cases.

\subsection{Hardware Overhead of Varying Random Bit Sizes}
To study the hardware overhead of different numbers of random bits, we again use the FP12 format without subnormal support.  The results, as presented in Table \ref{tab:random hardware}, show that even with a larger value of $r$, our SR implementation yields significant reductions in area, energy consumption, and latency compared to MAC units using half-precision formats. Putting it all together, our 12-bit SR design without support for subnormals reduces the delay, area and energy of the MAC unit by $\approx 50$\% w.r.t. FP32, while  maintaining near baseline accuracy during training. When compared to FP16, delay is reduced by more than $29$\%, and area and energy by $\approx 13$\%.

\begin{table}[t]
  \centering
  \caption{Impact of random bits $r$ on hardware overhead.}
  \label{tab:random hardware}
  \begin{tabular}{ccccccc}
    \hline
    Configuration                     & E                  & M                  & $r$  & Delay & Area        & Energy        \\
   &                    &                    &    & (ns)  & ($\mu m^2$) & ($\mu W/MHz$) \\

    \hline
    \multirow{5}{*}{SR eager W/O Sub} & \multirow{5}{*}{6} & \multirow{5}{*}{5} & 4  & 1.85  & 508.36      & 0.46          \\
    &                    &                    & 7 & 1.87  & 540.19      & 0.49          \\
    &                    &                    & 9 & 1.87  & 558.63      & 0.51          \\
    &                    &                    & 11 & 1.93  & 579.19     & 0.53          \\
    &                    &                    & 13 & \textbf{1.93}  & \textbf{601.71}      & \textbf{0,56}          \\
    \hline
    RN W/ Sub  (FP16)                       & 5                  & 10                 & -  & \textbf{2.73}  & \textbf{692.62}      & \textbf{0.65}          \\
RN W/ Sub (FP32)        & 8 & 23 & -  & \textbf{4.71}       & \textbf{1404.01}     & \textbf{1.17} \\
    \hline
  \end{tabular}
  \vspace{-10pt}
\end{table}


\section{Conclusion}

Using small number representations such as FP8 formats can significantly impact the computational demand of DNN training. However, in most of the state-of-the-art, accumulations are still done in either half (16-bit) or single (32-bit) precision arithmetic. While stochastic rounding in lower precision can be used to mitigate the loss of accuracy, it has traditionally come with hardware overhead and loss of performance.

In this paper we present a novel architecture for floating-point MAC units employing stochastic rounding. The proposed \textit{eager} design outperforms the classic \textit{lazy} one on all benchmarks, achieving up to 26.6\% latency and 18.5\% area savings. We also test our design in various DNN training settings, and by tuning the number of bits for the rounding unit we reach baseline-comparable results. 
Training results indicate that a configuration using 13 random bits and without subnormal support gives the best tradeoffs across all training benchmarks. Despite the hardware overhead due to more random bits, our design results in 29.3\% and 13.1\% savings in latency and area, respectively, w.r.t.~an FP16 accumulator with RN support. Furthermore, the hardware advantages of our proposed \textit{eager} design hold even greater potential within a systolic array-based accelerator, which we plan to cover in future work.


\bibliographystyle{IEEEtran}

\end{document}